\documentclass[a4paper,twoside]{article}

\usepackage{epsfig}
\usepackage{subcaption}
\usepackage{calc}
\usepackage{amssymb}
\usepackage{amstext}
\usepackage{amsmath}
\usepackage{amsthm}
\usepackage{multicol}
\usepackage{pslatex}
\usepackage{apalike}
\usepackage{algorithm2e}
\usepackage[bottom]{footmisc}

\usepackage{graphicx}% Include figure files
\usepackage{amsmath}
\usepackage{amssymb}
\usepackage{hyperref}
\usepackage{quantikz}
\usepackage{braket}
\usepackage{tikz}
\usepackage{lipsum}

\usepackage{orcidlink}
\usepackage{SCITEPRESS}     % Please add other packages that you may need BEFORE the SCITEPRESS.sty package.

\begin{document}

\title{Knapsack and Shortest Path Problems Generalizations From A Quantum-Inspired Tensor Network Perspective}

\author{\authorname{Sergio Muñiz Subiñas\sup{1}\orcidlink{0009-0008-7590-0149}, Jorge Martínez Martín\sup{1}\orcidlink{0009-0004-9336-3165}, Alejandro Mata Ali\sup{1}\orcidlink{0009-0006-7289-8827}, Javier Sedano\sup{1}\orcidlink{0000-0002-4191-8438} and Ángel Miguel García-Vico\sup{2}\orcidlink{0000-0003-1583-2128}}
\affiliation{\sup{1}Instituto Tecnológico de Castilla y León, Burgos, Spain}
\affiliation{\sup{2}Andalusian Research Institute in Data Science and Computational Intelligence (DaSCI), University of Jaén, 23071 Jaén, Spain}
\email{\{sergio.muniz, jorge.martinez, alejandro.mata, javier.sedano\}@itcl.es, agvico@ujaen.es}
}

\keywords{Quantum Computing, Tensor Networks, Combinatorial Optimization, Quantum-Inspired}

\abstract{In this paper, we present two tensor network quantum-inspired algorithms to solve the knapsack and the shortest path problems, and enables to solve some of its variations. These methods provide an exact equation which returns the optimal solution of the problems. As in other tensor network algorithms for combinatorial optimization problems, the method is based on imaginary time evolution and the implementation of restrictions in the tensor network. 
In addition, we introduce the use of symmetries and the reutilization of intermediate calculations, reducing the computational complexity for both problems.
To show the efficiency of our implementations, we carry out some performance experiments and compare the results with those obtained by other classical algorithms.}

\onecolumn \maketitle \normalsize \setcounter{footnote}{0} \vfill

\section{\uppercase{Introduction}}
Combinatorial optimization has become a wide field of study due to the large number of industrial and academic applications in different disciplines, for instance, engineering or logistics. Some of the most prominent and extensively studied problems are route optimization (traveling salesman problem~\cite{TSP,TSP2,TSP_Artificial,TSP_overview} or the shortest path~\cite{A-star,Dijkstra1959}), task scheduling (job-shop scheduling problem~\cite{Flexible_JSSP,JSSP_graph,JSSP_General}) and constrained optimization (knapsack~\cite{Knapsack_Google,Knapsack_original} or bin packing~\cite{Bin_Heuristic,Bin_Packing,Bin_Reinforcement}).
The fact that most of these problems are NP-hard makes exact resolution methods computationally unaffordable. With the objective of solving these problems, it is common to draw upon approximations, generally obtaining a less optimal result. However, in many industrial cases, the focus is on finding a solution that works well enough in a reasonable time, rather than looking for the best possible outcome. Some of the most used classical approximation techniques are greedy approaches~\cite{Greedy_TSP}, reinforcement learning~\cite{Bin_Reinforcement}, genetic algorithms~\cite{Genetic_FSTSP} or simulated annealing~\cite{simulated_annealing}, among others.

Meanwhile, due to the development of quantum technologies, new ways of approaching combinatorial optimization problems have been proposed. Among other applications, it is possible to address the solution for quadratic unconstrained binary optimization (QUBO)~\cite{QUBO_Tutorial} problems with quantum algorithms such as the quantum approximate optimization algorithm (QAOA)~\cite{QAOA} or the variational quantum eigensolver (VQE)~\cite{VQE}. Likewise, Constrained Quadratic Models (CQM) and Binary Quadratic Models (BQM) can be solved with quantum annealers~\cite{CQM_Quantum}. With the current state-of-the-art of quantum computers hardware, where qubits are noisy and unreliable, no improvement in the state-of-the-art has been achieved. Nevertheless, due to the strong theoretical basis of these algorithms, researchers have developed quantum-inspired strategies, with tensor networks emerging as one of the most populars~\cite{TN_Nut,Lectures_TN}. This discipline consists in taking advantage of some properties of tensor algebra to compress information or perform calculations in a more efficient way. Moreover, by applying tensor decomposition operations as singular value decomposition (SVD)~\cite{SVD} in a truncated way, it is possible to make approximations losing the least relevant information. Numerous studies show the computational advantages of tensor networks in a wide variety of applications, for instance, in the simulation of quantum systems~\cite{Lectures_TN}, compression of neuronal networks~\cite{TN_NN_Compresion}, anomaly detection~\cite{Anomaly_Detection,Anomaly_detection_tensortrain} or combinatorial optimization problems~\cite{Combinatorial_Optimization}, which are the object of study of this paper. 
 
In this paper we present two tensor network algorithms to solve the knapsack and shortest path problems. Our method is based on previous works applied to traveling salesman problem~\cite{TSP_TN}, QUBO problems~\cite{Qudo_Qubo} and task scheduling~\cite{Task_Scheduling_TN}. Following the results in~\cite{melocoton}, it provides an exact equation that returns the optimal solution to these problems. The objective of this study is to present an analytical methodology for solving combinatorial optimization problems, rather than a computational advantage in such problems. We compare the results of our implementations against some state-of-the-art algorithms. Finally, we present more complicated generalizations of these two problems, which are hard to solve, and how to solve them easily with the tensor network method.

The main contributions of this work are:
\begin{itemize}
    \item Providing an exact and explicit equation to solve knapsack and shortest path problems.
    \item Introducing a quantum-inspired algorithm to solve them on a classical computer.
    \item Presenting additional algorithms for novel versions of these problems.
\end{itemize}

This paper is structured in the following way. First, we analyze the knapsack problem in Sec.~\ref{sec: knapsack}, introducing the problem, the tensor network algorithm, the contraction scheme and the computational complexity. Then we analyze in the same way the shortest path problem in Sec.~\ref{sec: shortest}. After that, we analyze our methods performance and compare it with classical optimizers in Sec.~\ref{sec: results}. Finally, we address generalizations for these problems in Sec.~\ref{sec: generalizations}.

\section{\uppercase{Knapsack Problem}}\label{sec: knapsack}

Given a set of $n$ different item classes, where the $i$-th class can be selected up to $c_i$ times, each with an associated weight $w_i\in\mathbb{N}$ and value $v_i\in\mathbb{R}^+$, the knapsack problem consists in finding the configuration $\vec{x}=(x_0,..,x_{n-1})$ that maximizes
\begin{equation}\label{eq: Value equation}
    \begin{gathered}
        V(\vec{x})=\sum_{i=0}^{n-1}x_iv_i \\
        \text{subject to }W(\vec{x})=\sum_{i=0}^{n-1}x_iw_i\leq{Q},\\
        x_i\in [0,c_i]\quad \forall i\in [0,n-1],
    \end{gathered}
\end{equation}
being $x_i$ the number of times the $i$-th class has been selected, $V(\vec{x})$ the total value of the items, $W(\vec{x})$ the total weight of the items in the knapsack and $Q$ the maximum weight capacity.

There are different variants of this problem, the most popular being the 0-1 knapsack, where each class can be selected only once, $x_i\in\{0,1\}$. This case can be interpreted as an analogous way to formulate the general one, creating a class for each time an item can be selected. Another version is the unbounded knapsack problem, where each class can be selected without bounds except for the maximum weight capacity $x_i\in\mathbb{N}$. All these instances can be solved exactly with a time complexity of $\mathcal{O}(nQ)$ and space complexity $\mathcal{O}(nQ)$.

In addition, there are more complicated variants in which multiple variables need to be maximized or additional constraints have to be satisfied. However, the optimal solution is not needed in most of the real-world applications, so it is possible to use algorithms that have less computational complexity at the cost of finding a less optimal solution.

\subsection{Tensor Network Algorithm} \label{KP Tensor Network algorithm}
In this section we describe the quantum-inspired tensor network used, how the algorithm works, and some important details related to the implementation. To address the knapsack problem, we need to identify the configuration of $\vec{x}$ that maximizes the value $V(\vec{x})$. 
The idea of our algorithm is to take advantage of the imaginary time evolution to generate a state whose basis state with the greatest amplitude represents the optimal configuration. Due to the high number of possible combinations, we cannot know the amplitude of all combinations at once, and hence determine the highest probability state. Therefore, we employ a method that consists in determining each variable separately, by projecting the state into each subspace associated with the respective variable.
\subsubsection{Quantum-Inspired Tensor Network}
\begin{figure}
    \centering
    \includegraphics[width=0.9\linewidth]{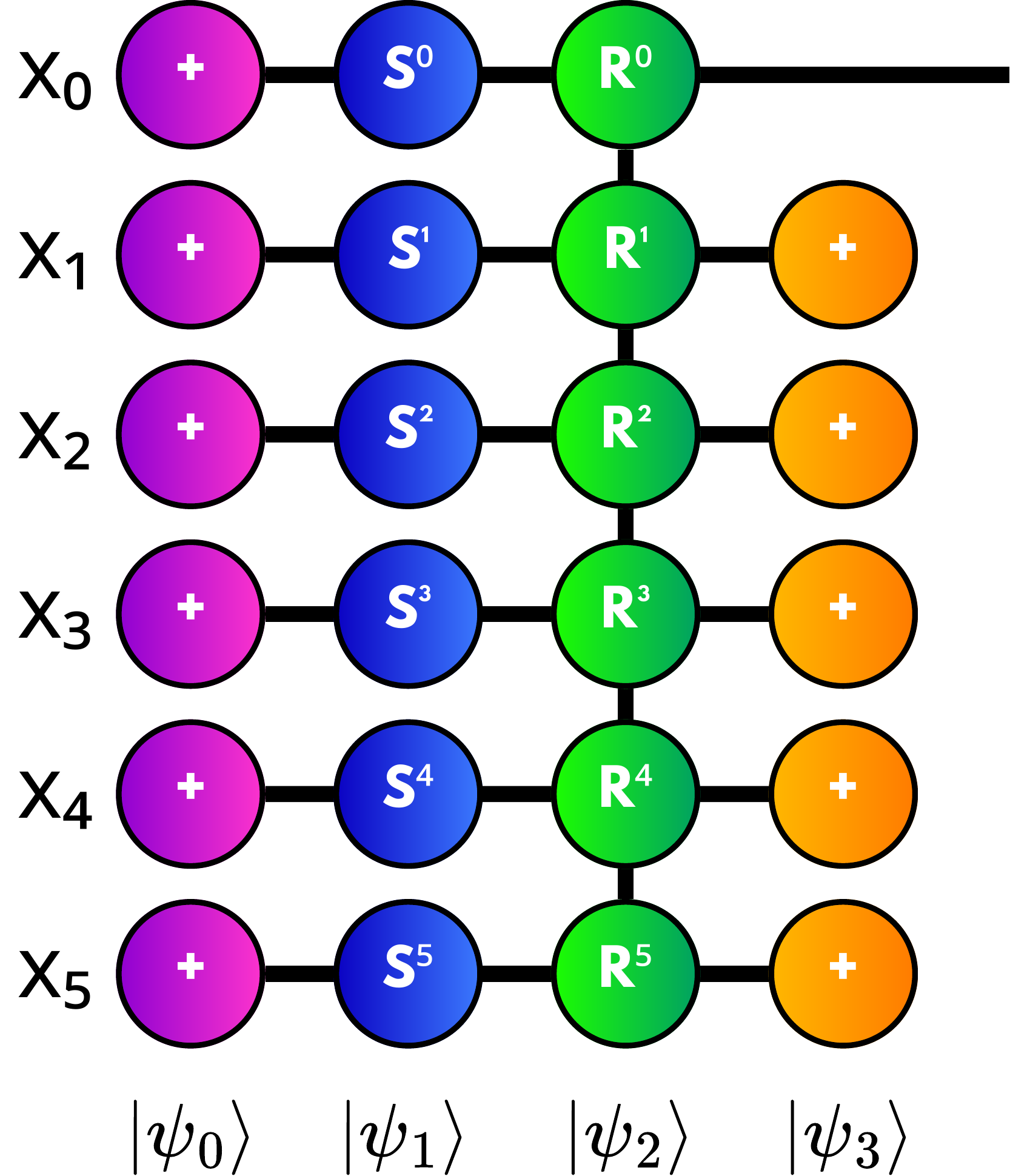}
    \caption{Tensor network solving the knapsack problem composed of four column layers. From left to right: superposition (`+'), evolution ($S$), constraint ($R$) and tracing (`+').}
    \label{fig: knapsack network}
\end{figure}
Our knapsack problem algorithm is based on tensor algebra and some properties of quantum computing such as superposition or projected measurements. Thus, we have developed a quantum-inspired tensor network that incorporates the theoretical advantages of quantum computing without being limited by its constraints. The tensor network (see Fig. \ref{fig: knapsack network}) can be interpreted as a `quantum circuit' that can be simulated in four steps, each one corresponding to one of the layers in the figure.

The system starts in a uniform superposition of $n$ variables, each one encoded in the basis states of a qudit of dimension $c_i+1$. This first step corresponds to the first layer of tensors `+' in Fig.~\ref{fig: knapsack network}, and the mathematical representation of the resulting state is
\begin{equation}\label{eq: superposition}
    |\psi_0\rangle = \sum_{\vec{x}}|\vec{x}\rangle=\bigotimes_{i=0}^{n-1}\sum_{x_i=0}^{c_i}|x_i\rangle.
\end{equation}
The next step consists of applying an imaginary time evolution layer to the superposition that amplifies the basis states with a higher value. We implement this with an operator $\mathcal{S}$, which assigns to each state $\ket{\vec{x}}$ an exponential amplitude corresponding to its associated value $V(\vec{x})$. This is integrated within the tensor network as the union between the $i$-th qudit with a diagonal evolution tensor $S^i$, which multiplies each $\ket{x_i}$ basis state by $e^{\tau v_i x_i}$. The parameter $\tau$ scales the costs so that the amplitudes differ exponentially from each other. The state after applying this layer is
\begin{align}
    |\psi_1\rangle =& \mathcal{S}|\psi_0\rangle= \bigotimes_{i=0}^{n-1}\sum_{x_i=0}^{c_i}S^i|x_i\rangle =\nonumber\\
    =& \bigotimes_{i=0}^{n-1}\sum_{x_i=0}^{c_i}e^{\tau{v_i}x_i}|x_i\rangle = \sum_{\vec{x}}e^{\tau{V(\vec{x}})}|\vec{x}\rangle.
\end{align}
In order to eliminate incompatible configurations, we apply a third layer that imposes the maximum capacity constraint,
\begin{equation}
    \begin{gathered}
    |\psi_2\rangle = \mathcal{R}(Q)|\psi_1\rangle=\sum_{\vec{x}}e^{\tau{V(\vec{x})}}\mathcal{R}(Q)|\vec{x}\rangle \\
    \ \mathcal{R}(Q)\ket{\vec{x}} =\ket{\vec{x}} \text{ if } W(\vec{x})\leq{Q},
    \end{gathered}
\end{equation}
where $\mathcal{R}(Q)$ is the operator that projects the state into the subspace that fulfills the maximum weight capacity $Q$. This is implemented by a layer of $R^i$ tensors, where the up index keeps track of the current weight loaded at the knapsack until that qudit, while the down index outputs the resultant weight after adding the new qudit state. If this weight exceeds $Q$, the amplitude of the state is multiplied by zero.

Now that we have a system that has performed the minimization and constraint, we need a method to extract the position of the largest element of the superposition without the need of checking all the possible combinations. To do this, we have to take into account that, for a sufficiently large $\tau$, there will be a very high amplitude peak in the state of the optimal solution. If we want to determine the first variable, we can sum the amplitudes of all the states that have $0$ as their value for the first variable, all those with $1$, all those with $2$, etc. In this way, if the peak is high enough, the sum that has the correct value for the first variable will have such a large sum that it will be greater than all the other sums. Thus, we only need to add over the possible outcomes of the other qudits for each possible outcome of the first, and determine that the correct value is the one with the highest associated sum. This process is similar to measure one of the qubits in the quantum system.

In order to achieve this, we apply a partial trace layer to the tensor network except for the first qudit. This leads to the state
\begin{align}
\ket{\psi_3^0} 
 =&\sum_{\vec{z}}\langle z_1,\dots,z_{n-1}|\psi_2\rangle =\nonumber\\ =&\sum_{\vec{z}}\sum_{\vec{x}}e^{\tau{V(\vec{x}})}\langle z_1,\dots,z_{n-1}| \mathcal{R}(Q)|\vec{x}\rangle = \nonumber \\ 
    =& \sum_{\vec{x}|W(\vec{x})\leq{Q}}e^{\tau{V(\vec{x}})}|x_0\rangle. 
\end{align}
It can be interpreted as the resulting state from projecting $|\psi_2\rangle$ into the Hilbert space $\mathcal{H}^{c_0+1}$ of the first qudit. The associated vector representation is:
\begin{equation}
\ket{\psi_3^0}=\begin{pmatrix}
\braket{0+\dots+|\psi_2} \\
\braket{1+\dots+|\psi_2} \\
\vdots\\
\braket{c_0+\dots+|\psi_2}
\end{pmatrix},
\end{equation}
where each amplitude allows to determine the probability of $\ket{\psi_3^0}$ of being in their corresponding basis states $\ket{x_0}.$

\begin{figure}
    \centering
    \includegraphics[width=0.9\linewidth]{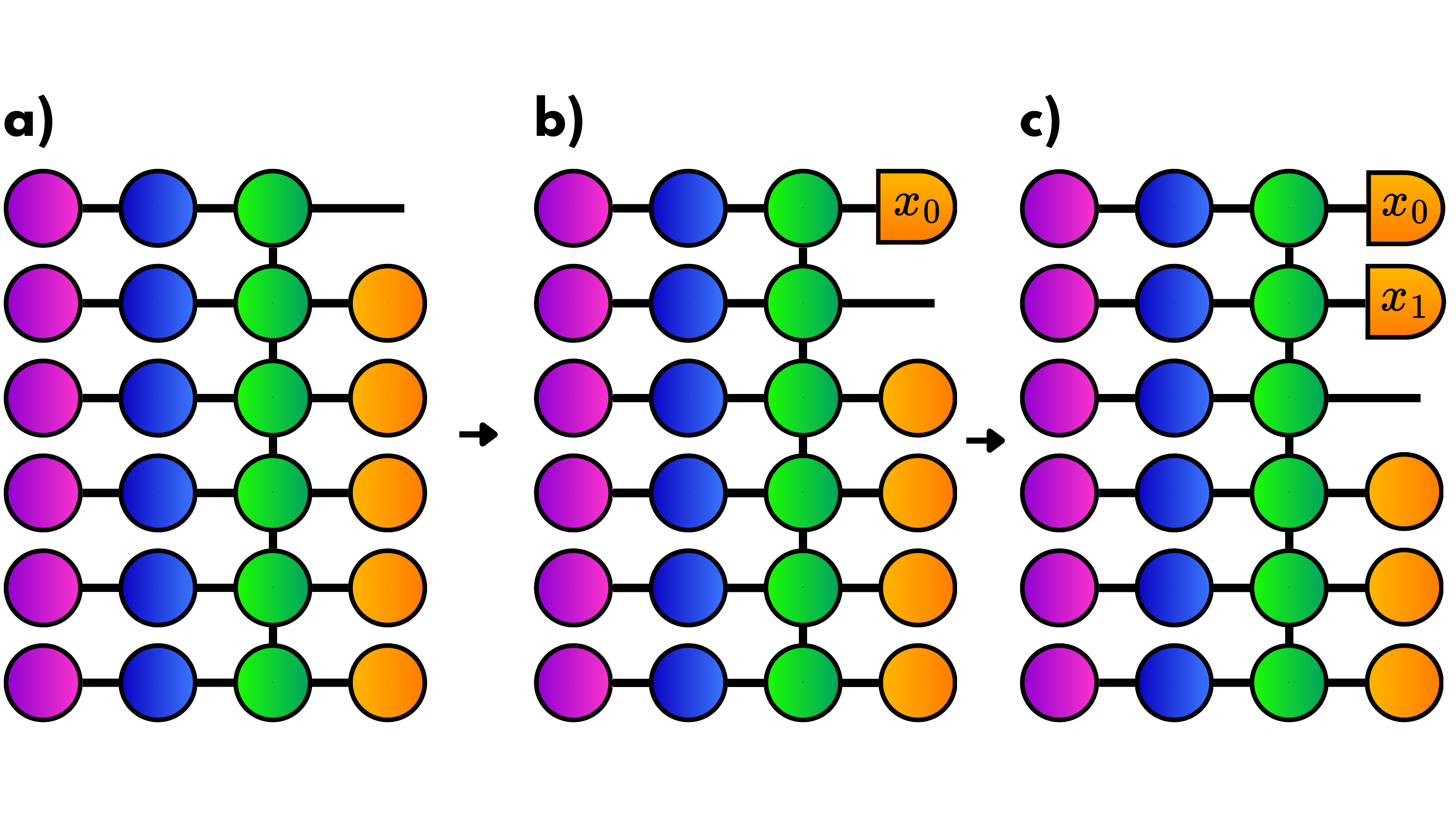}
    \caption{Graphic representation of the iterative method of the tensor network to obtain each variable.}
    \label{fig: variable proyection}
\end{figure}

To determine the first variable $x_0$, we contract the tensor network (see Fig. \ref{fig: knapsack network}) to obtain $\ket{\psi_3^0}$, whose highest amplitude value position corresponds to the optimum value of the variable $x_0$. To compute the remaining variables, we use the same tensor network but, instead of leaving free the index associated to the first variable, we free the index we want to obtain and follow the same contraction process as in the previous step. With this, we obtain the $\ket{\psi^m_3}$ state, from which we determine the $m$-th variable. In each step, at the end of the line of the variables that we have already determined, we put a vector where all its elements are zero, except the element corresponding to the chosen value of the variable. This process is shown in Fig. \ref{fig: variable proyection}. 

Additionally, we can take advantage of the results we have already obtained in order to reduce the solution space and the contraction complexity. To achieve this, once we have determined the value of a qudit, we can build a new tensor network that excludes the row of that qudit while incorporating the information from the measurement of the removed qudit. This gives the same result because the information already determined is classical and fixed, so it does not need any superposition.

\subsubsection{Tensor Network Layout}
\begin{figure}
    \centering
    \includegraphics[width=0.9\linewidth]{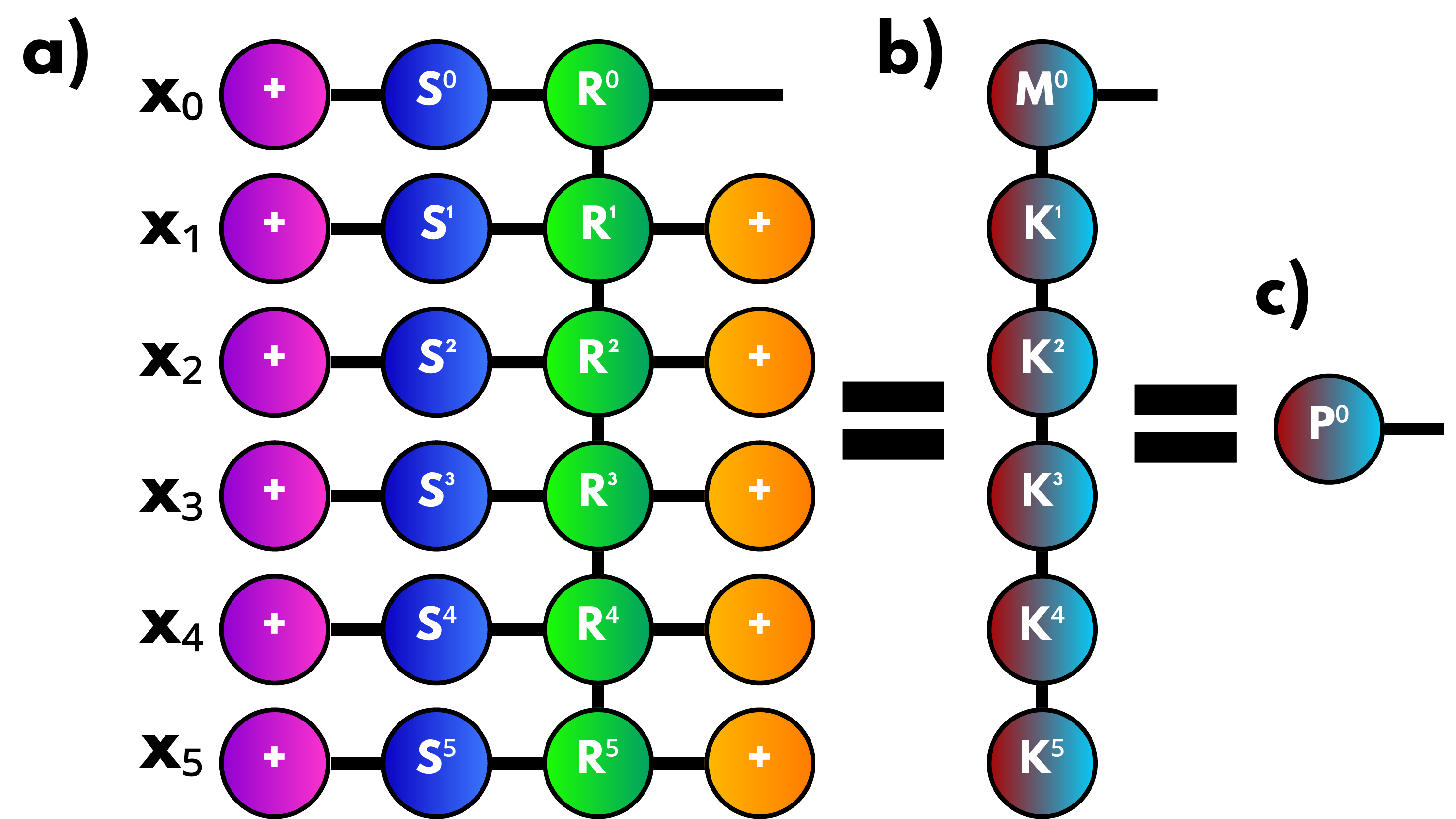}
    \caption{Tensor network solving the knapsack problem. a) Extended version. b) Compressed version into a linear TN that implements all four layers simultaneously. c) Contracted tensor.}
    \label{fig: contraction scheme}
\end{figure}
Since it is possible to compress the information of the four layers into a single layer in a convenient way, we propose a direct implementation of the tensor network in a chain structure (see Fig. \ref{fig: contraction scheme}). We define the tensors following the index convention of Fig. \ref{fig: tensors indexes} and the notation used to define the tensors mathematically can be found in the appendix.
\begin{figure}
    \centering
    \includegraphics[width=0.9\linewidth]{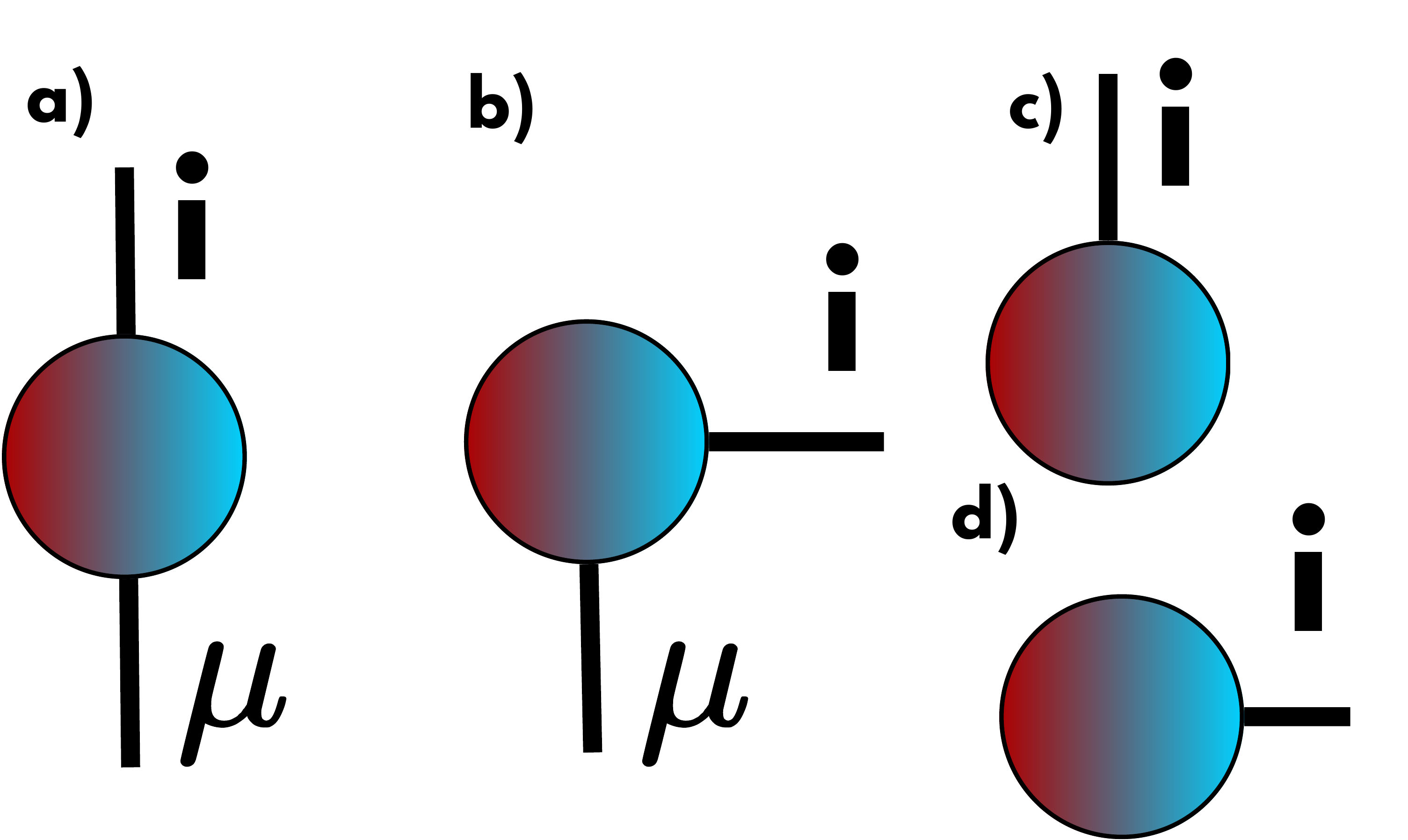}
    \caption{Nomenclature of the indexes of each tensor from the tensor chain, where $i$ corresponds to a variable index and $\mu$ to a dependent index. a) A tensor from the middle of the tensor chain. b) The top tensor of the tensor chain. c) The bottom tensor of the tensor chain. d) The resulting tensor from the contraction of the tensor chain.}
    \label{fig: tensors indexes}
\end{figure}

From the figure, we can see that there are three different types of tensors. The tensor $M^0$, with elements $M^0_{i\mu}$, is associated with the first qudit we want to measure. This tensor is the result of the contraction between the superposition node `+', the evolution node $S^0$ and the projector node $R^0$. It outputs information about the state of the qudit throughout the index $i$, and outputs the weight of the knapsack after adding the new items throughout the index $\mu$. Therefore, the dimensions of the indexes $i$ and $\mu$ are $c'_0$ and $Q'$ respectively, being $Q'=Q+1$ and $c'_i=c_i+1$ to streamline the notation. The non-zero elements of the matrix $M^0_{c'_0 \times Q'}$ are defined as
\begin{align}
\mu &= iw_0  \nonumber\\ 
M^0_{i\mu} &= e^{\tau i v_0}.
\end{align}

The following $n-2$ tensors $K^k$ belong to the same type, they have elements $K^k_{i\mu}$ and are associated with the $k$-th qudit. Each of these tensors receives information about the weight of the knapsack until the previous qudit throughout the index $i$, and passes the weight of the knapsack, including the weight added by this qudit, throughout the index $\mu$. Therefore, the dimensions of the indexes $i$ and $\mu$ are both $Q'$. The non-zero elements of $K^k_{Q' \times Q'}$ are defined as 
\begin{equation}
    \begin{gathered}
    y_k \in [0,c_k], \\
    \mu=i+y_k w_k, \\
    K^{k}_{i\mu}=e^{\tau y_k v_k},
    \end{gathered}
\end{equation}
where $y_k$ is the number of times we introduce the $k$-th element into the knapsack. This considers all possible additions of elements of that class that do not exceed the maximum weight.

Finally, the last tensor $K^{n-1}$, with elements $K^{n-1}_{i}$, is associated with the last qudit and receives from $i$ the information of the knapsack of all the other qudits. The non-zero elements of $K^{n-1}_{Q'}$ are 
\begin{equation}
    \begin{gathered}
    y_{i}=\min\left(\frac{Q-i}{w_{n-1}},c_{n-1}\right), \\
    K^{n-1}_{i}=e^{\tau y_i v_{n-1}}, 
    \end{gathered}
\end{equation}
being $y_i$ the maximum amount of elements of the class $n-1$ that could be added without exceeding the maximum weight capacity $Q$. Since in the event that one can have up to $y_i$ positive value elements, the most optimal is always to have the maximum number.

It is important to note that by cutting the chain as the values of the variables are being set, the upper tensor of the chain will be different in each iteration, as it must include the information from the results obtained in the previous steps. Thus, during the iteration where the variable $x_m$  is determined, our tensor network consists of the same $n-m-1$ last tensors and a new upper tensor $M^{m}_{c_m'\times Q'}$
\begin{equation}
    \begin{gathered}
    \mu=i w_{m} + \sum_{k=0}^{m-1}x_k w_k, \\
    M^{m}_{i\mu}=e^{\tau i v_{m}}.
    \end{gathered}
\end{equation}

All these tensors can be optimized in dimensions, having in mind the maximum weight that can be reached up to the variable associated to such tensor. However, this does not change the computational complexity, remaining only as an implementation aspect.

It is interesting to note that the provided tensor network is the exact formula that solves the combinatorial problem. This allows us to analyze the problem from a different perspective at the mathematical level, although we will not do so in this paper.

\subsection{Contraction Scheme}\label{ssec: contraction}

In order to optimize the use of computational resources, we develop a contraction scheme that takes advantage of intermediate calculations. The idea is to store the resulting vectors from each step of the contraction process, while contracting the tensor chain bottom-up. This way, for each iteration of the algorithm, we only need to contract the new initial tensor with the corresponding stored intermediate tensor.
\begin{figure}
    \centering
    \includegraphics[width=0.9\linewidth]{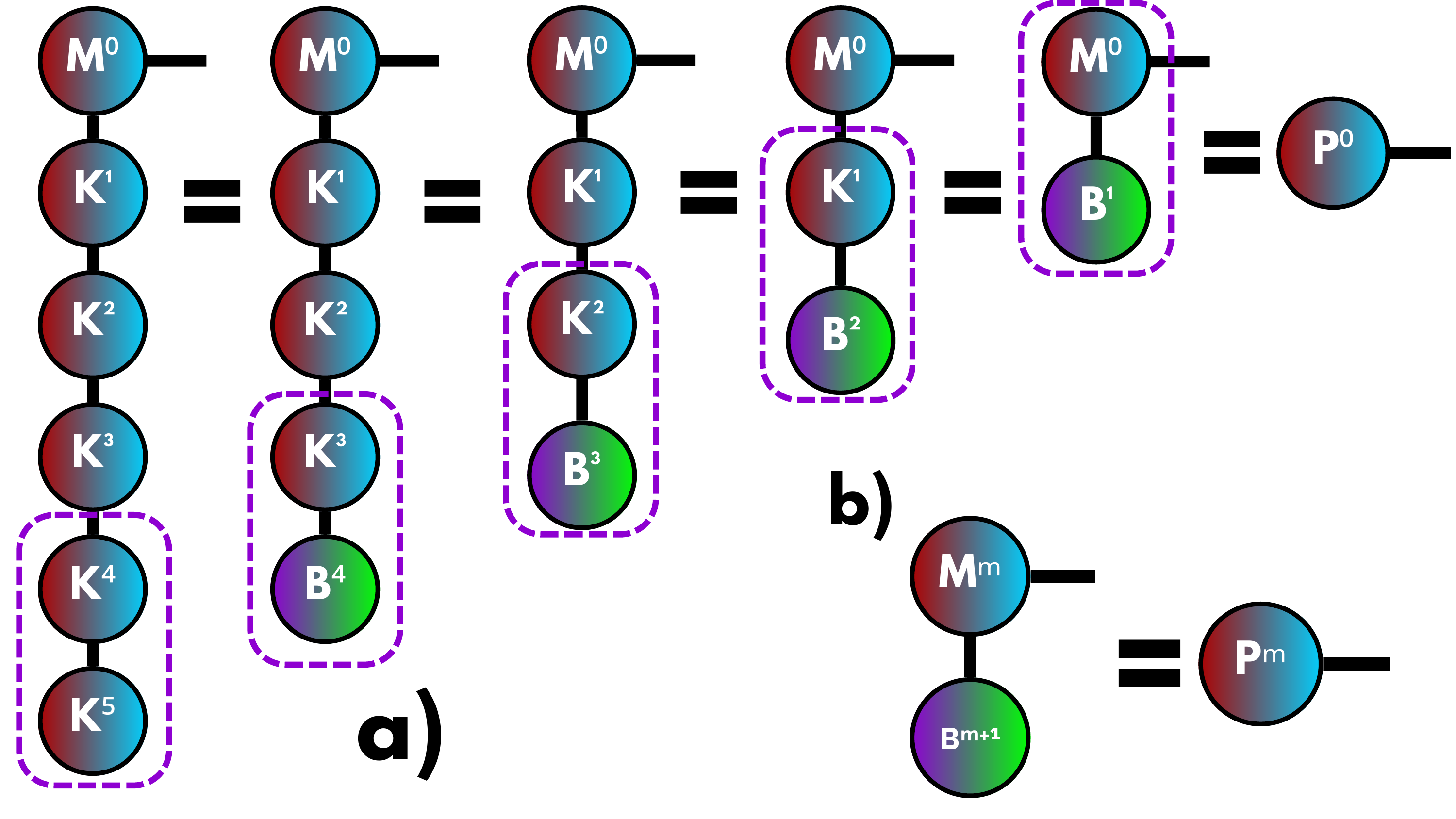}
    \caption{Contraction scheme with storage of intermediate tensors for later calculations.}
    \label{fig: contraction_scheme}
\end{figure}

The implementation of this recursive method can be seen in Fig. \ref{fig: contraction_scheme}. In this process, the last node of the chain is contracted with the node directly above, resulting in a new node $B^k$, which will be stored in memory for later operations. We have to repeat this process until we have a single node $P^0$ which represents the $\ket{\psi^0_3}$ state. Once all the tensors are stored, we only need to contract the $M^m$ matrices with their corresponding $B^{m+1}$ tensor to obtain $\ket{\psi^m_3}$, contracting the whole tensor network of each iteration in one step.

\subsection{Optimizations and Computational Complexity}
As we have mentioned earlier, the tensor network algorithm that we implement needs to store about $n$ $Q\times Q$ matrices. Therefore, to carry out the proposed contraction scheme, we need to store the result of $n$ matrix-vector operations, in addition to performing each contraction between $M^m$ and $B^{m+1}$.

Since each vector-matrix contraction has complexity $\mathcal{O}(Q^2)$, the computational cost of the first step is $\mathcal{O}(nQ^2)$, which is not optimal considering the structure of the matrices. It is possible to verify that the tensors $K^k$ present sparsity; in particular, the number of nonzero elements in these matrices depends on $c_k$. All elements of $K^k$ are zero except for $c_k+1$ diagonals, each one starts in column $c_k^i w_k$ where $c_k^i \in [0, c_k]$ and all its elements are equal to $e^{c_k^i\tau v_k}$. Therefore, the number of non-zero elements is $\mathcal{O}(Qc_k)$.

Keeping this in mind and knowing that the computational complexity of multiplying a dense vector with a sparse matrix is $\mathcal{O}(n_z)$, where $n_z$ is the number of non-zero elements, the computational complexity of this step is $\mathcal{O}(nQc)$, with $c$ being the average $c_k$. As the second step also involves the multiplication of a sparse matrix and a dense vector, the computational complexity is likewise of the order
$\mathcal{O}(nQc)$. Thus, the total time complexity of the algorithm is $\mathcal{O}(nQc)$, which is $\mathcal{O}(nQ)$ in the 0-1 case.

For space complexity we have to take into account that each intermediate vector stored has $\mathcal{O}(Q)$ elements. The matrices can be generated dynamically while contracting, so they require only $\mathcal{O}(c)$ elements each, which is negligible compared to $\mathcal{O}(Q)$. The total space complexity is $\mathcal{O}(nQ)$.

In the case where intermediate calculations are not used, the contraction process must be carried out each time a variable is to be determined. This approach does not have the problem of storing $n$ vectors of dimension $Q$, being the space complexity $\mathcal{O}(Q)$. However, the computational complexity becomes $\mathcal{O}(n^2Qc)$.

\section{\uppercase{Shortest path problem}}\label{sec: shortest}

Given a graph $G$ with $V\in\mathbb{N}$ vertices and a set $E$ of edges, the shortest path problem, also called the single-pair shortest path problem, consist in finding a $n$-step path between two vertices
\begin{equation}
\begin{gathered}
    \vec{v}=(v_0,v_1,...,v_{n-1}),\\
    v_t \in [0,V],
\end{gathered}
\end{equation}
 such us the cost of the route is minimized, where $v_t$ is the vertex associated to the $t$-th step. The cost of the route is given by
\begin{equation}
    \begin{gathered}
    C(\vec{v})=\sum_{t=0}^{n-2}E_{v_t,v_{t+1}}, \\
    \end{gathered}
\end{equation}
 where $E_{ij}\in\mathbb{R^+}$ corresponds to the cost between the $i$-th vertex and $j$-th vertex. If two vertices are not connected, the cost between them is $E_{ij}=\infty$, and the cost between a node and itself is $E_{ii}=0$.
 
There are small variations of the problem depending on the directionality of the graph, where it could be either directed or undirected, or the weights of the edges, which could be weighted or unweighted.
 
All previous variations can be generalized into a directional weighted graph. In the case of the unidirectional graph, we just set $E_{ij} = E_{ji}$, and in the case of the unweighted graph, we just change the cost of the edges to one.
 
The single-pair directional shortest path can be solved with a time complexity of $\mathcal{O}((E+V)\log(V))$ and a space complexity $\mathcal{O}(V)$~\cite{Dijkstra_Complexity}.

\subsection{Tensor Network Algorithm}
\begin{figure}
    \centering
    \includegraphics[width=0.7\linewidth]{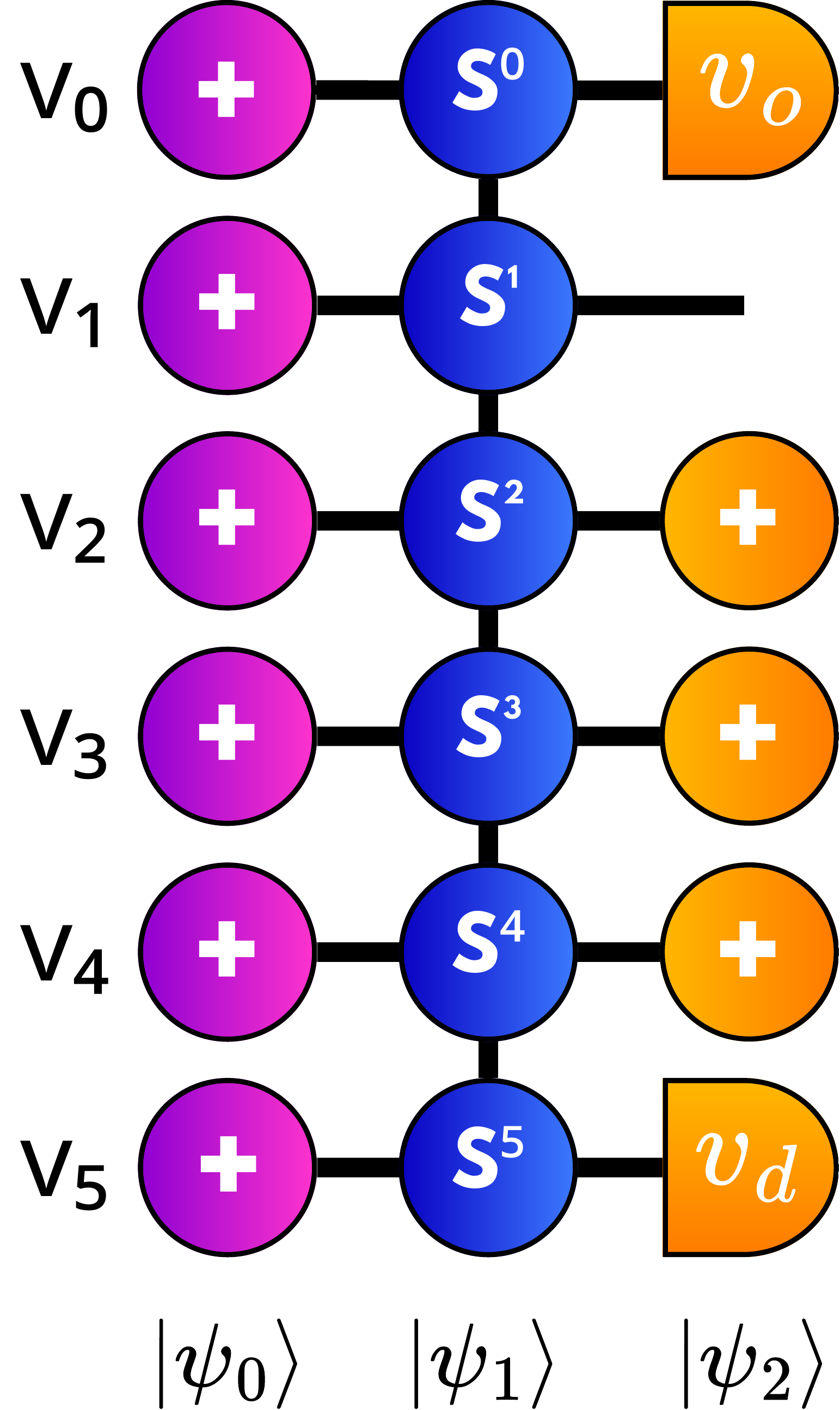}
    \caption{Tensor network solving the shortest path problem composed of three column layers. From left to right: superposition (`+'), evolution ($S$) and tracing (`+').}
    \label{fig: Shortest path tensor network layout}
\end{figure}
The algorithm we propose in this paper is based on the tensor network in Fig. \ref{fig: Shortest path tensor network layout}, relaying on the same principles explained in Ssec. \ref{KP Tensor Network algorithm}.

\subsubsection{Quantum-Inspired Tensor Network}

The tensor network starts with a superposition layer equivalent to the state in Eq. \ref{eq: superposition}.
In this superposition, we apply an imaginary time evolution layer that damps the basis states with higher-cost routes. The state after applying this layer is
\begin{equation}
    |\psi_1\rangle = \sum_{\vec{v}}e^{-\tau{C(\vec{v}})}|\vec{v}\rangle=\bigotimes_{t=0}^{n-2}\sum_{v_t=0}^{V-1}e^{-\tau E{v_t,v_{t+1}}}|v_t\rangle.
\end{equation}

The last layer consists of a partial trace layer in each qudit except the second. For the first and last nodes, it is necessary to connect them to a node that represents the origin and destination nodes respectively, instead of the superposition node. This leads to the state

\begin{align}
\ket{\psi_2^1} =&\sum_{\vec{z}}\langle v_o,z_2,\dots,v_d|\psi_1\rangle =\nonumber\\ =&\sum_{\vec{z}}\sum_{\vec{v}}e^{-\tau{C(\vec{v}})}\langle v_o,z_2,\dots,v_d|\vec{v}\rangle = \nonumber \\ 
    =& \sum_{\vec{v}}e^{-\tau{C(\vec{v}})}|v_1\rangle, 
\end{align}
where $v_o$ corresponds to the origin node and $v_d$ to the destiny node.
\begin{figure}
    \centering
    \includegraphics[width=1\linewidth]{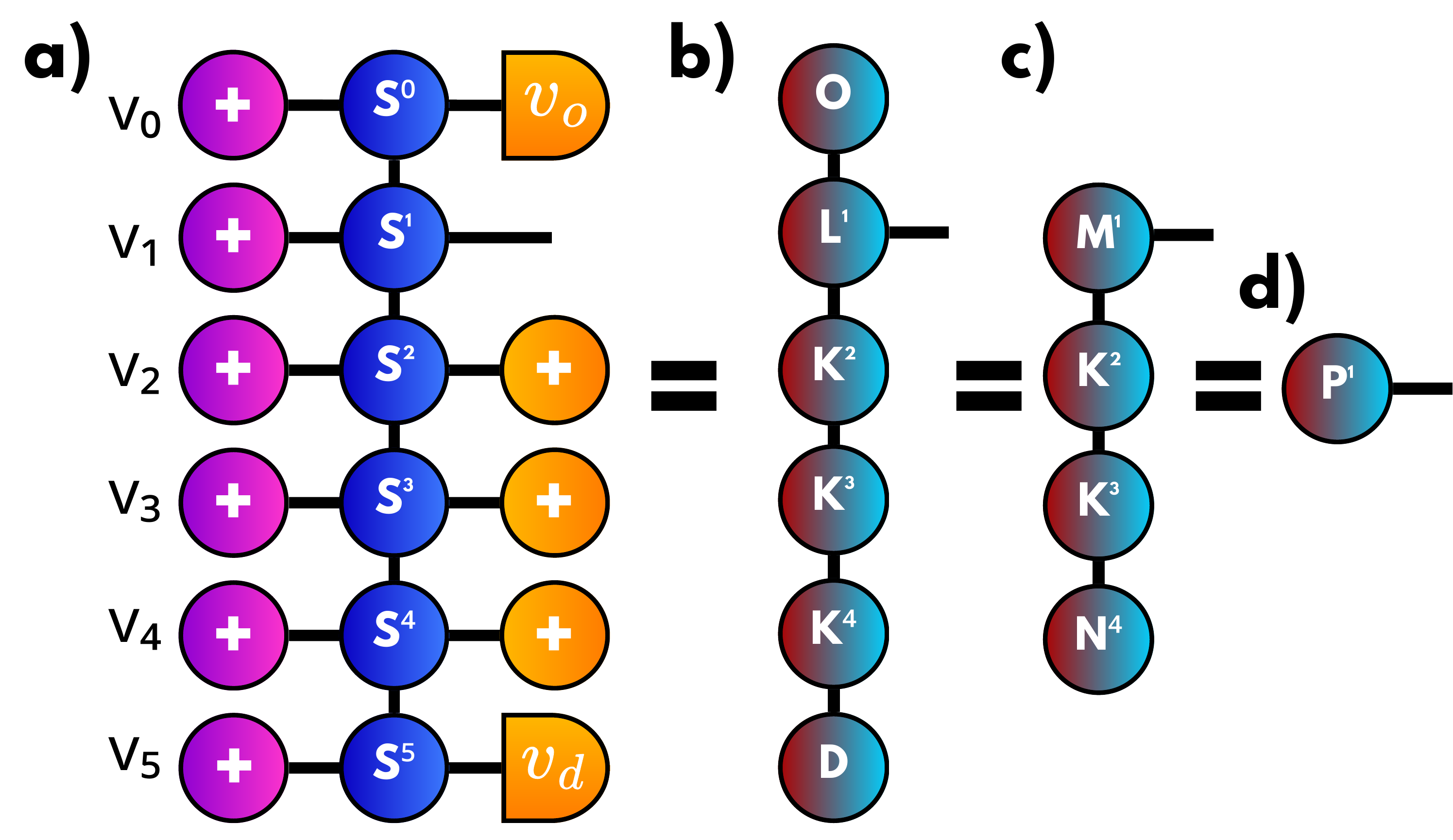}
    \caption{Tensor network solving the shortest path problem. a) Extended version. b) Compressed version into a linear TN that implements all three layers simultaneously. c) Compressed version with the fixed nodes removed. d) Contracted tensor.}
    \label{fig: Shortest path contration scheme}
\end{figure}

\subsubsection{Tensor Network Layout}
As in the case of the knapsack, we propose a compressed formulation of the tensor network in a chain structure (see Fig. \ref{fig: Shortest path contration scheme}). 

From the figure, we can see that there are three different types of tensors. The tensor $L^1$, with elements $L^1_{ijk}$, is associated with the qudit we want to measure. Given that the origin node is fixed, we remove it by manually adding the information to the next node resulting in the node $M^1_{ij}$. This tensor is the result of the contraction between the superposition node `+', the evolution node $S^1$ and the trace node. It outputs information about the state of the qudit throughout the index $i$, and outputs the selected node throughout the index $j$. Therefore, the dimensions of the indexes $i$ and $j$ are both $V$. The matrix $M^1_{V\times V}$ is defined as
\begin{equation}
M^1_{ij} = e^{-\tau (E_{oi}+E_{ij})}.
\end{equation}

The following $n-3$ tensors belong to the same type, the tensor $K^k_{V \times V}$ which has elements $K^k_{ij}$ is associated with the $k$-th qudit. It receives information about the previous visited node throughout the index $i$, and passes the new visited node throughout the index $j$. Therefore, the dimension of the indexes $i$ and $j$ is $V$  for both of them. $K^k_{V\times V}$ is defined as 
\begin{equation}
    K^{k}_{ij} = e^{-\tau E_{ij}}.
\end{equation}

Finally, the last non-fixed tensor $K^{n-2}_{ij}$ is associated with the second, since the last node corresponds to the destiny node. It receives from $i$ the information of the previous node and, by fixing the last node to the destiny, we add the cost of the edge between the second last and the last node removing the index $j$. This result in the node $N^{n-2}_{V}$ with the non-zero elements being 
\begin{equation}
    N^{n-2}_{i}=\sum_{j=0}^{V-1}e^{-\tau (E_{ij} + E_{jd})}.
\end{equation}

It is important to note that by cutting the chain as the values of the variables are being set, the upper tensor of the chain will be different in each iteration as it must include the information from the results obtained in the previous steps. Thus, during the iteration where the variable $v_m$  is determined, our tensor network consists of the same $n-m-1$ last tensors and a new upper tensor $M^{m}_{ij}$
\begin{equation}
    \begin{gathered}
    M^{m}_{ij}= e^{-\tau (E_{pi}+E_{ij})},
    \end{gathered}
\end{equation}
where $p$ corresponds to the previous visited node.
This tensor network follows the same contraction scheme as explained in Ssec. \ref{ssec: contraction}.

\subsection{Optimizations and Computational Complexity}
As mentioned before, the tensor network algorithm that we implement needs to store about $n$ $V\times V$ matrices. Therefore, to carry out the proposed contraction scheme, we need to store the result of $n$ matrix-vector operations, in addition to performing the contractions between $M^m$ and $B^{m+1}$. As each vector-matrix contraction has complexity $\mathcal{O}(E)$, the computational cost of the first step is $\mathcal{O}(nE)$. Each following step has a complexity of $\mathcal{O}(E)$, so that the total complexity is $\mathcal{O}(nE)$.

Given that the $n$ tensors are equal except for the last, we only need to store one copy of the connectivity matrix to perform all operations. Although we need to store the result of the vector-matrix contractions, it can be stored with a complexity of $\mathcal{O}(V^2)$ or, considering that each non zero element corresponds to an edge, it can also be defined as $\mathcal{O}(E)$.  

With the reuse of intermediate steps we only need to store $n$ vectors of dimension $V$, so the total space complexity is $\mathcal{O}(nV+E)$.

\section{\uppercase{Results and Comparatives}}\label{sec: results}
After explaining the theoretical framework of the algorithms, we present a series of experiments to compare our approaches with the state-of-the-art. We evaluate both, the quality of the solutions versus $\tau$ and the execution time versus the size of the problem.

\subsection{Knapsack Results}
The original idea was to compare our algorithm with the Google Or-tools solver. However, due to its limitation of requiring $v_i$ to be integer and its significantly lower maximum weight capacity $Q$, we have excluded this option. Therefore, the algorithm used for the comparisons is a greedy approach, which is capable of finding a good approximate solution in almost no time for every problem instance. Basically, it consists of arranging the items in the knapsack in ascending order according to the value/weight ratio, and then inserting objects until the knapsack is filled to the maximum.

\begin{figure}
    \centering
    \includegraphics[width=1\linewidth]{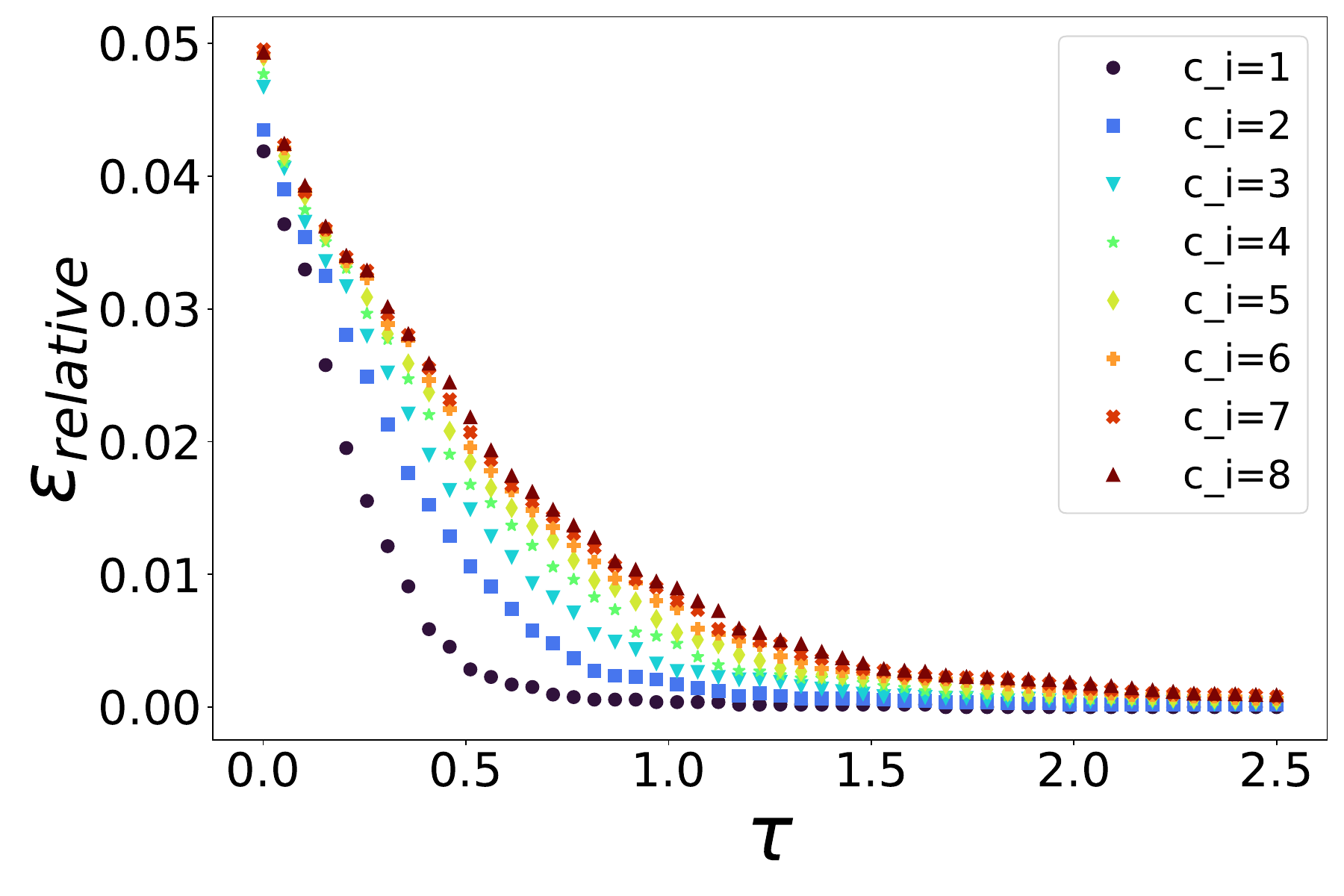}
    \caption{Relative error $\epsilon_{relative}= 1-V(\vec{x}_{tn})/V(\vec{x}_{gr})$ depending on $\tau$ and $c_i$. Each problem has 1000 classes and a capacity of $0.8\sum_{i=0}^{n-1} c_i w_i$.}
    \label{fig: precition-tau}
\end{figure}

To demonstrate that our implementation works, we perform an experiment to show that the algorithm obtains results as good as the greedy one. Fig. \ref{fig: precition-tau} displays how the algorithm performs as a function of $\tau$ and the number of elements in all classes (in this case, all classes have the same number of elements). Our indicator is the relative error defined as
\begin{equation}
    \epsilon_{relative} = 1-V(\vec{x}_{tn})/V(\vec{x}_{gr}),
\end{equation}
being $V(\vec{x}_{tn})$ the value of the tensor network solution and $V(\vec{x}_{gr})$ the value of the greedy solution. As expected, as $\tau$ increases, the relative error $\epsilon_{relative}$ decreases, causing the solution to approximate the greedy approach.

To fairly compare the results for the different cases of $c_i$, the values of each class have been divided by $c_i$. Thus, it is found that the case $c_i=1$ requires a smaller $\tau$ to converge to the optimal solution. This is expected because the number of possible combinations grows as $c_i$ increases.

It should be noted that when $\tau$ reaches a certain value, the product of the exponential may result in an overflow, causing the algorithm to produce suboptimal solutions. This implies that, although this algorithm is theoretically an exact method, in practice it is not possible to solve arbitrarily large problems. When exponential saturation is reached, the algorithm begins to perform worse and, for this reason, it is crucial to select a value $\tau$ that is large enough to distinguish the optimal solution while remaining small enough to prevent saturation.

\begin{figure}
    \centering
    \includegraphics[width=1\linewidth]{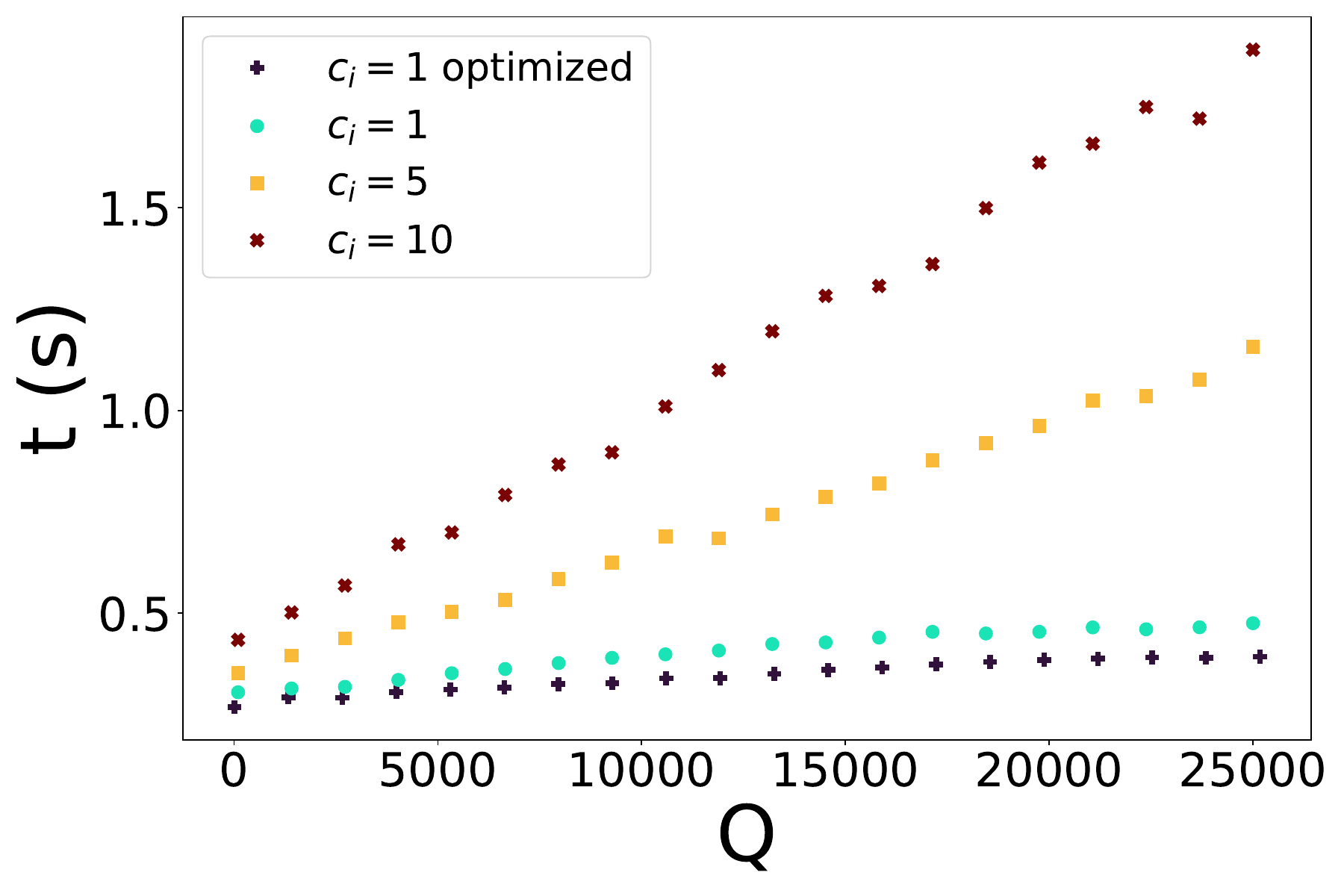}
    \caption{Dependency between the capacity of the knapsack $Q$ and the execution time for different problem instances. Use of intermediate calculations, number of classes $n = 5000$, $\tau = 1$.}
    \label{fig: capacity-time}
\end{figure}

\begin{figure}
    \centering
    \includegraphics[width=1\linewidth]{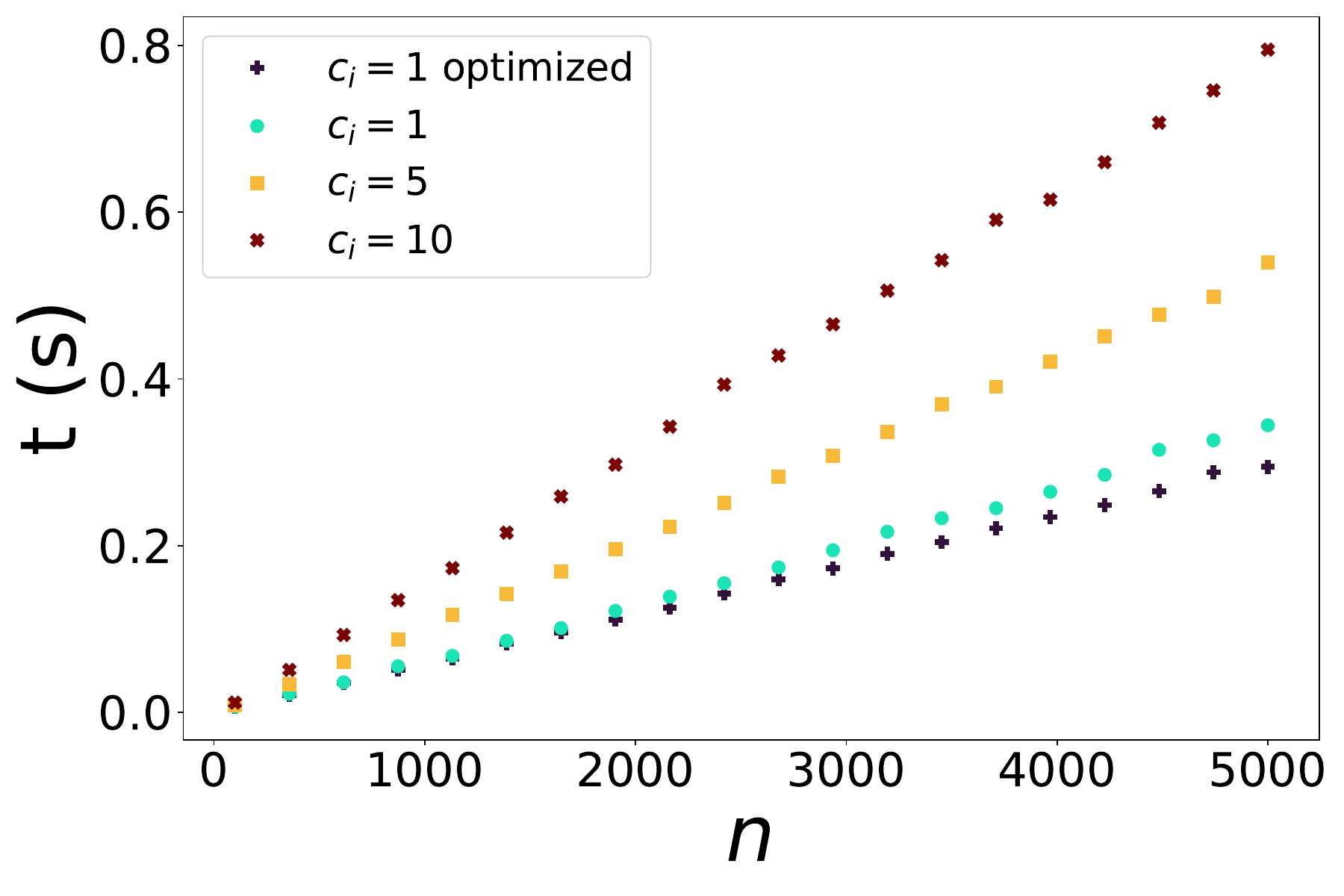}
    \caption{Dependency between the number of classes $n$ and the execution time for different problem instances. Use of intermediate calculations, maximum weight capacity $Q = 7000$, $\tau = 1$.}
    \label{fig: clases-time}
\end{figure}

\begin{figure}
    \centering
    \includegraphics[width=1\linewidth]{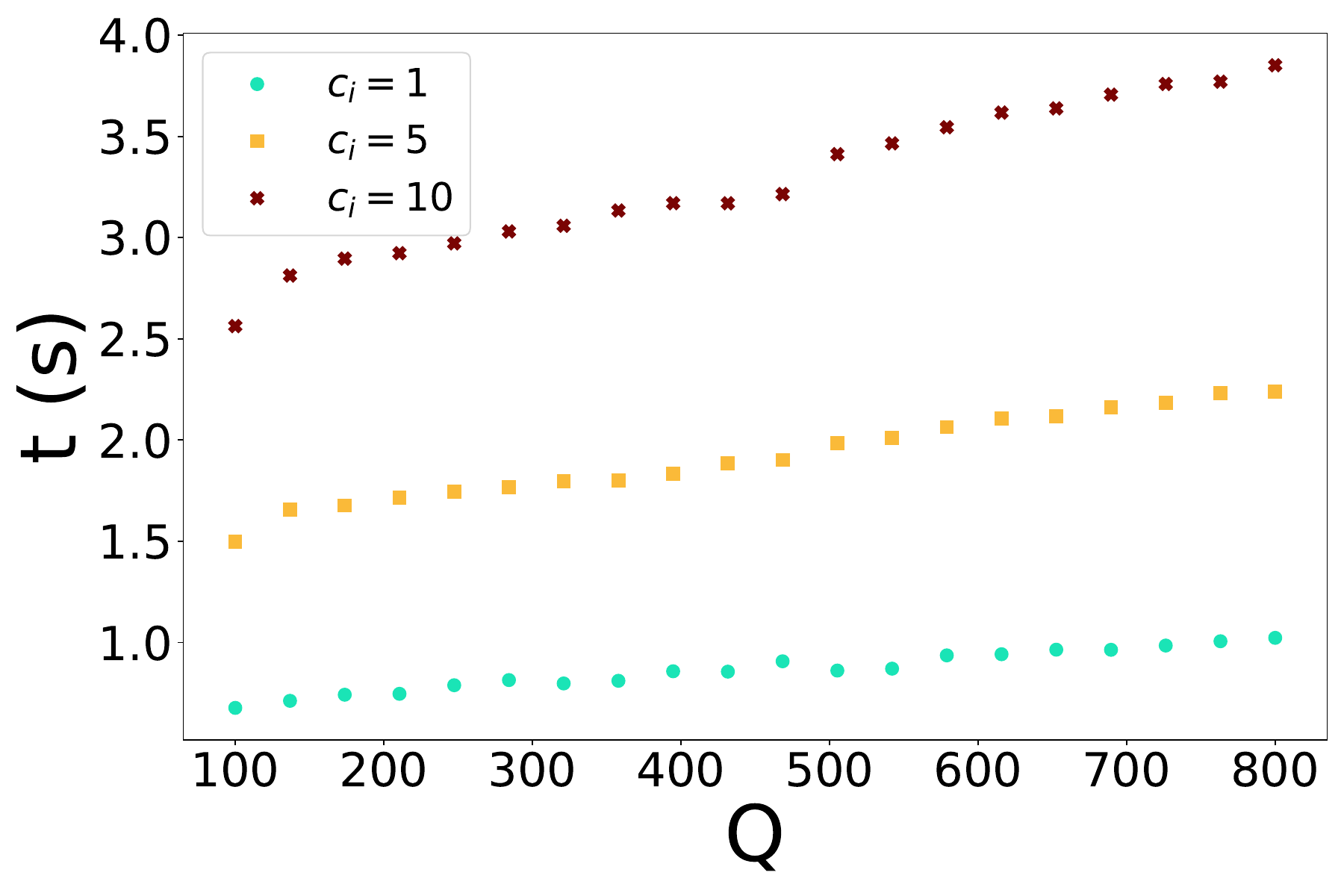}
    \caption{Dependency between the capacity of the knapsack $Q$ and the execution time for different problem instances. No use of intermediate calculations, number of classes $n = 500$, $\tau = 1$.}
    \label{fig: capacidad-time_sincalinter}
\end{figure}

\begin{figure}
    \centering
    \includegraphics[width=1\linewidth]{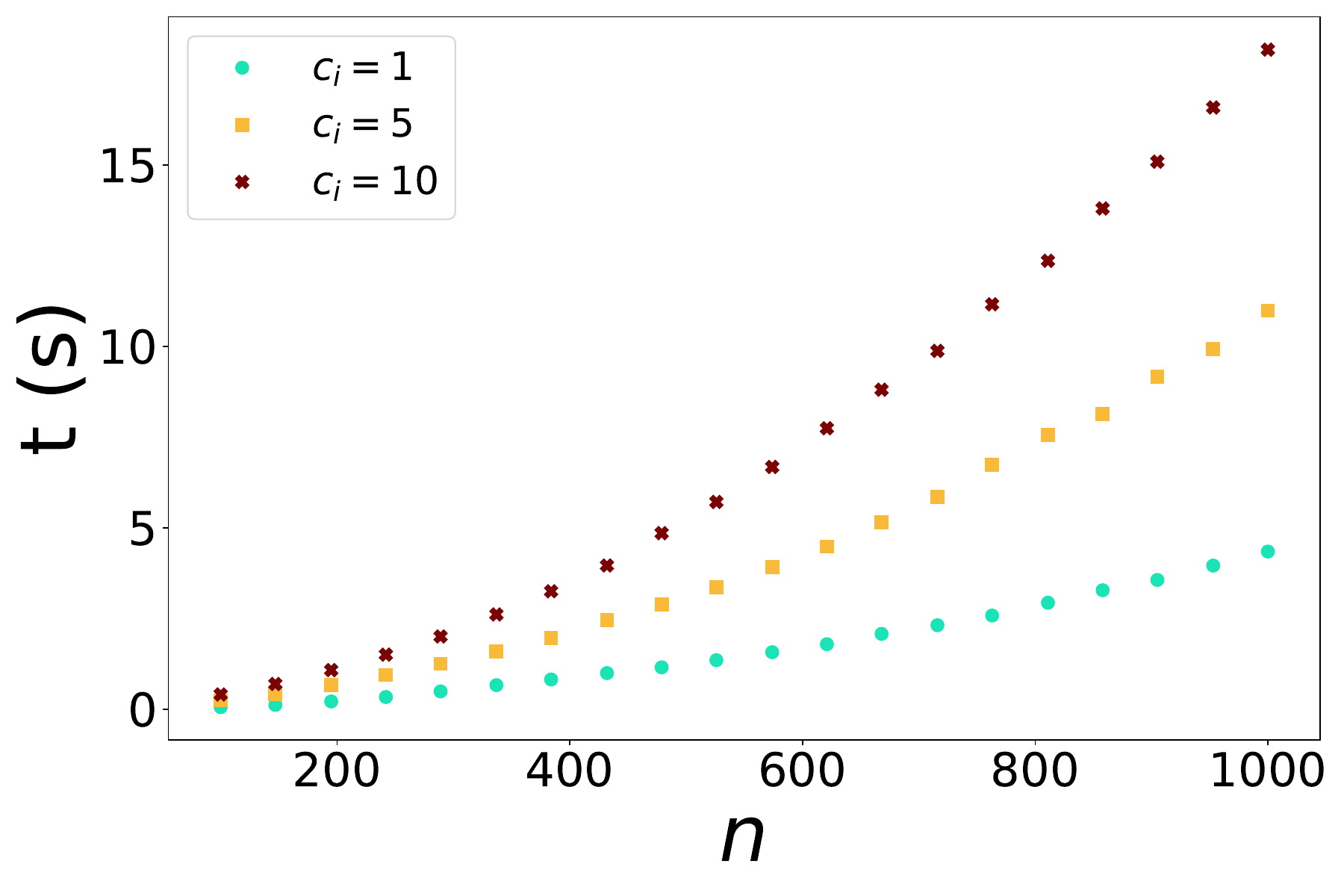}
    \caption{Dependency between the number of classes $n$ and the execution time for different problem instances. No use of intermediate calculations, maximum weight capacity $Q = 1500$, $\tau = 1$.}
    \label{fig: clases-time_sincalinter}
\end{figure}

Additionally, as mentioned in the theoretical section, it is possible to verify in Figs. \ref{fig: capacity-time} and \ref{fig: clases-time} that by taking advantage of intermediate calculations, the computational complexity in time of the algorithm depends linearly with the maximum weight capacity $Q$ and the number of classes $n$. If intermediate calculations are not used, there is a linear dependency in time versus $Q$ and a quadratic in time versus $n$ (Figs. \ref{fig: capacidad-time_sincalinter} and \ref{fig: clases-time_sincalinter}).


After performing all the tests, we determine that our algorithm is able to solve knapsack problems with $n=1000$ variables and small $c_i$ values. However, for larger instances, the process begins to overflow, leading to less optimal solutions. With the disadvantage of longer computational time, it is possible to achieve better results using the Decimal library. In fact, for problems with up to 1000 classes, our algorithm is able to find better solutions than greedy in several problem instances.

\subsection{Shortest Path Problem}
\begin{figure}
    \centering
    \includegraphics[width=1\linewidth]{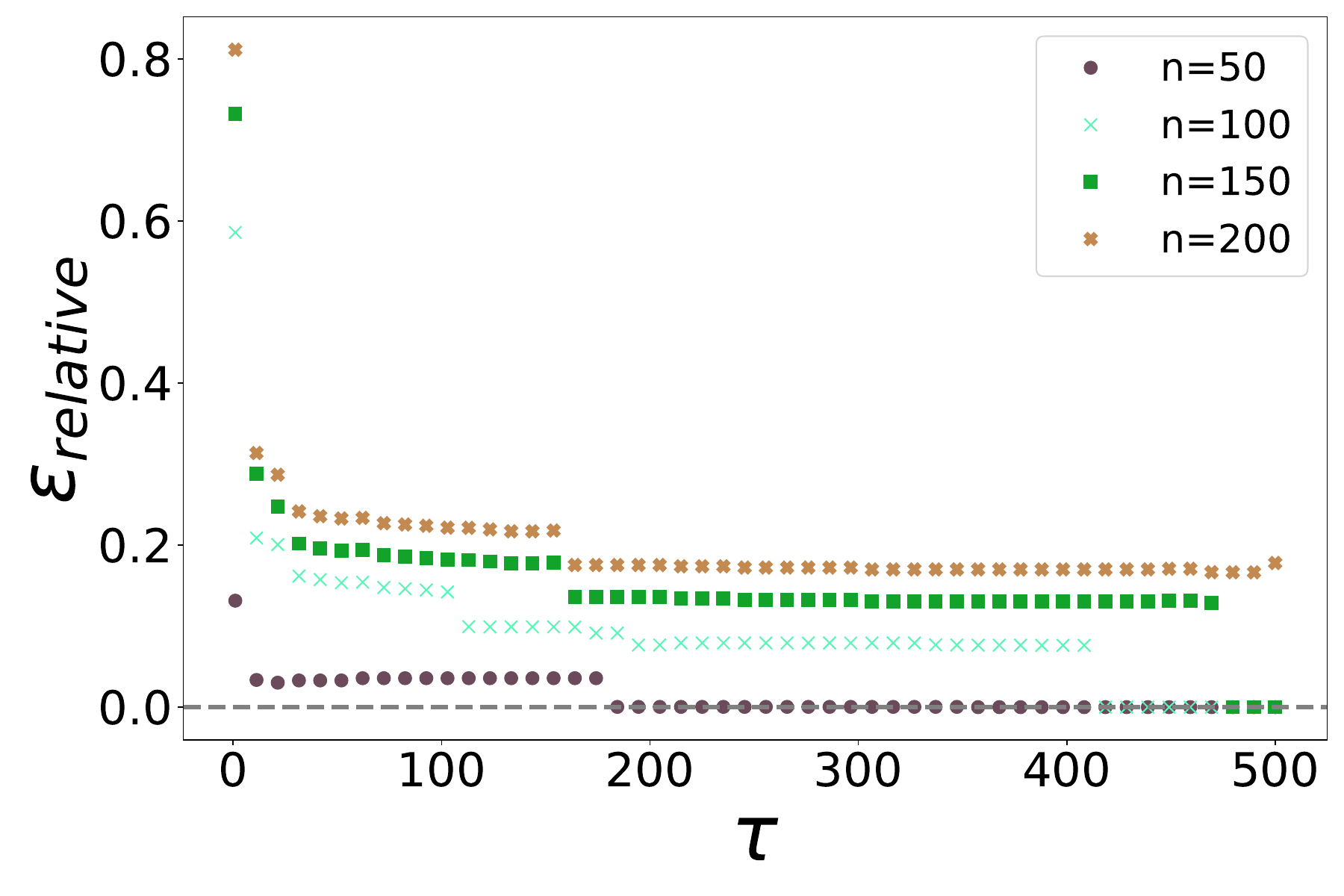}
    \caption{Relative error $\epsilon_{relative}=C(\vec{x}_{tn})/C(\vec{x}_{exact})-1$ depending on $\tau$ and $n$. Each problem has $V=12408$.}
    \label{fig: error-tau-shortest}
\end{figure}

We compare our implementation with the Dijkstra's algorithm~\cite{Dijkstra1959}. In order to compare how our algorithm performs against Dijkstra's, we propose an experiment where both algorithms have to go from the same origin node to the same destiny node within the same graph. This graph corresponds to Valladolid city (Spain). Given that our algorithm needs to have a fixed number of steps in order to work, we compare how the results evolve according to $\tau$ and the number of steps. As can be seen in Fig. \ref{fig: error-tau-shortest}, both the number of steps and the value of $\tau$ influence the result. The higher the number of total steps, the greater $\tau$ needs to be in order to reach the optimal solution. Our indicator is the relative error defined as
\begin{equation}
    \epsilon_{relative}=C(\vec{x}_{tn})/C(\vec{x}_{exact})-1,
\end{equation}
 being $C(\vec{x}_{tn})$ the value given by the tensor network and $C(\vec{x}_{exact})$ the number given by Dijkstra's algorithm. The relative error $\epsilon_{relative}$ decreases as $\tau$ increases, bringing the solution closer to the optimal.

\begin{figure}
    \centering
    \includegraphics[width=1\linewidth]{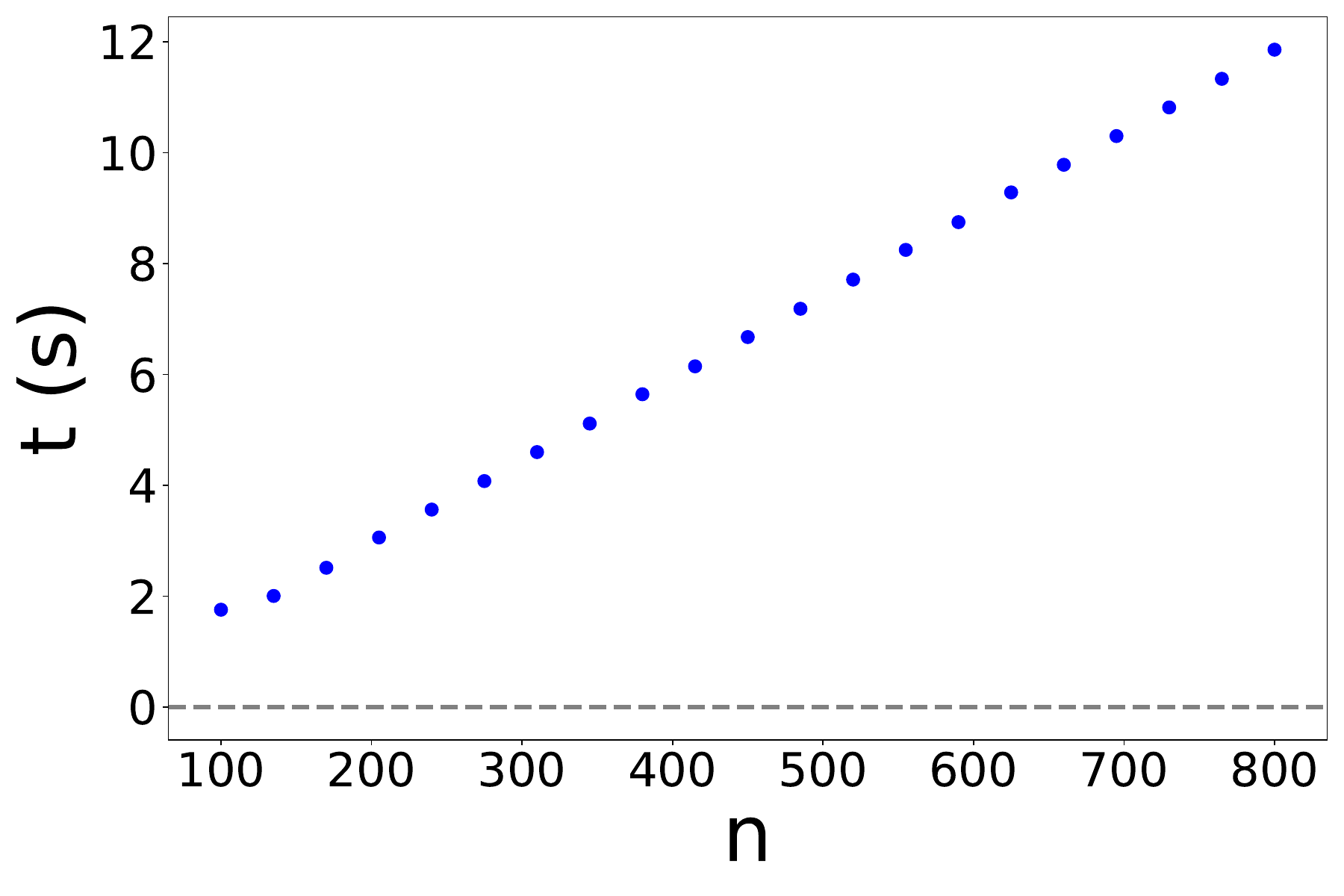}
    \caption{Execution time for different number of steps with $V=12408$ and $\tau=600$ for Valladolid, Spain.}
    \label{fig: time-steps}
\end{figure}

In addition, we compare the number of steps with the execution time. As shown in Fig. \ref{fig: time-steps}, the execution time scales linearly with the number of steps.

Since the number of steps is the most relevant parameter when it comes to finding the right path, in Fig. \ref{fig: time-steps} we have represented the relation between $n$ and the execution time of the algorithm and as can be seen they are linearly related.

\section{\uppercase{Generalizations}}\label{sec: generalizations}
In this section, we propose several generalizations of these problems and how to modify our tensor network algorithms to easily address them.
\subsection{Non Linear Knapsack Problem}
An interesting generalization of the knapsack problem is the nonlinear case. Under these circumstances, our cost and weight functions are given by the sum of nonlinear functions such that we define it as
\begin{equation}
    \begin{gathered}
        V(\vec{x})=\sum_{i=0}^{n-1}v_i(x_i) \\
        \text{subject to }W(\vec{x})=\sum_{i=0}^{n-1}w_i(x_i)\leq{Q},\\
        x_i\in [0,c_i]\quad \forall i\in [0,n-1],
    \end{gathered}
\end{equation}
where $w_i$ and $v_i$ are functions that receive natural numbers and return natural and real positive respectively.

Another simpler formulation is to convert these functions into vectors, since their inputs are natural, so that the problem is rewritten as
\begin{equation}
    \begin{gathered}
        V(\vec{x})=\sum_{i=0}^{n-1}v_{i,x_i} \\
        \text{subject to }W_{\vec{x}}=\sum_{i=0}^{n-1}w_{i,x_i}\leq{Q},\\
        x_i\in [0,c_i]\quad \forall i\in [0,n-1],
    \end{gathered}
\end{equation}
where $w$ is a natural number tensor with $w_{i,b}=\infty$ when $b>c_i$ and $v$ is a positive real number tensor with $v_{i,b}=-\infty$ when $b>c_i$ .

To address this problem, we will only need to modify the exponentials of the imaginary time evolution to add the nonlinearity and modify the output rates of the weight outputs. In this way, the tensors $M^m_{c'_m\times Q'}$ will have their non-zero elements
\begin{equation}
    \begin{gathered}
    \mu=w_{m,i} + \sum_{k=0}^{m-1}w_{k,x_k}, \\
    M^{m}_{i\mu}=e^{\tau v_{m,i}},
    \end{gathered}
\end{equation}
the tensors $K^k_{Q'\times Q'}$ will have their non-zero elements
\begin{equation}
    \begin{gathered}
    y_k \in [0,c_k], \\
    \mu=i+w_{k,y_k}, \\
    K^{k}_{i\mu}=e^{\tau v_{k,y_k}},
    \end{gathered}
\end{equation}
and the tensors $K^{n-1}_{Q'}$  will have their non-zero elements
\begin{equation}
    \begin{gathered}
    d_{i}=\arg\max(\vec{\rho^{i}}),\\
    \rho^{i}_{y} = \frac{1}{Q-i-w_{n-1,y}}, \\
    K^{n-1}_{i}=e^{\tau v_{n-1,d_i}}, 
    \end{gathered}
\end{equation}
where $d_i$ is the maximum number of elements of the last class that can be introduced into the knapsack without exceeding the capacity $Q$ having already a weight $i$.

In this case, the tensors involved are of the same size as in the original case, and we can also use the reuse of intermediate calculations in the same way. In addition, the optimization of the diagonals works exactly the same, only now they will not be equispaced. For all this, the computational complexity of the algorithm is the same as in the original case.

\subsection{Polynomial Knapsack Problem}
Another important generalization of the knapsack problem is the case where the weight function is a polynomial. To generalize as much as possible, we will take the value function to be given as a sum of nonlinear functions, and the weight function is a polynomial of a sum of nonlinear functions. With the tensorial notation, the problem is expressed as
\begin{equation}
    \begin{gathered}
        V(\vec{x})=\sum_{i=0}^{n-1}v_{i,x_i}, \\
        W_{\vec{x}}=\sum_{i=0}^{n-1}w_{i,x_i},\\
        F(z)=a_0+a_1z+a_2z^2+\dots+a_pz^p\\
        \text{subject to }F(W_{\vec{x}})\leq{Q},\\
        x_i\in [0,c_i]\quad \forall i\in [0,n-1],
    \end{gathered}
\end{equation}
where $w$ is a natural number tensor with $w_{i,b}=\infty$ when $b>c_i$ and $v$ is a positive real number tensor with $v_{i,b}=-\infty$ when $b>c_i$ .

In this case, what the tensors will send to each other will be the partial result of $W$ up to that point, exactly as in the previous case. However, the change will be in the last tensor, which will be the one that will apply the $F$ function on the total accumulated $W$, eliminating the states for which it exceeds $Q$. The last tensor is
\begin{equation}
    \begin{gathered}
    d_{i}=\arg\max(\vec{\rho^{i}}),\\
    \rho^{i}_{y} = \frac{1}{Q-F(i+w_{n-1,y})}, \\
    K^{n-1}_{i}=e^{\tau v_{n-1,d_i}}, 
    \end{gathered}
\end{equation}
where $d_i$ is the maximum number of elements of the last class that can be introduced into the knapsack without exceeding the capacity $Q$ having already a weight $i$.

Given the characteristics of this modification, we can also address the case in which $F$ is not a polynomial of $z$, but is a nonlinear function of $z$. In this case, the modification of the final tensor is exactly the same, taking into account the new $F$ function.

If we impose that the coefficients $a_k$ are positive integers, then $Q'\geq F(W_{\vec{x}})\geq W_{\vec{x}}$, so the matrices will have, at most, dimension $Q'$.

For the same reasons as in the previous case, the computational complexity is the same as in the original case.

\subsection{Time Dependent Shortest Path Problem}
This generalization consists of having a shortest path problem in which the network changes every time we take a step. That is, after each trip between nodes, the network changes in a predefined way, both in connectivity and in travel costs. This problem is interesting for cases such as tourist trip planning. The cost of each trip between nodes is the cost of travel between locations and the cost of hotel stays, as they change between days.

The mathematical modeling now is to look for the route that minimizes cost
\begin{equation}
    \begin{gathered}
    C(\vec{v})=\sum_{t=0}^{n-2}E^t_{v_t,v_{t+1}}, \\
    \end{gathered}
\end{equation}
where each cost also depends on the time step.

The way to solve this problem with our tensor networks method is to take a modified version of the algorithm presented in Sec. \ref{sec: shortest}. Instead of each $K^k$ tensor being equal, each one has some nonzero elements following
\begin{equation}
    K^{k}_{ij} = e^{-\tau E^k_{ij}}.
\end{equation}
The first and last tensors now are
\begin{equation}
M^m_{ij} = e^{-\tau \left(E^{m-1}_{x_{m-1}i}+E^m_{ij}\right)}.
\end{equation}
\begin{equation}
    \begin{gathered}
    K^{n-2}_{i}=\sum_{j=0}^{V-1}e^{-\tau \left(E^{n-2}_{ij} + E^{n-1}_{jd}\right)},
    \end{gathered}
\end{equation}

Because we have to contract the same number of matrices with the same size and shape, being able to reuse intermediate calculations, the computational complexity in time is the same. However, because we have $n$ time steps in which the $K$ matrices are different, the space complexity is multiplied by $n$, so that we have $\mathcal{O}(nE)$.

\section{\uppercase{Conclusions}}

This work introduces a quantum-inspired tensor network approach to solving the knapsack and the shortest path problems. The methodology is based on some quantum computing properties, such as superposition, entanglement, or quantum measurements. Theoretically, the proposed algorithms allow to determine the solutions of the problems exactly. However, due to the overflow caused by the implementation, the efficiency of the algorithm is limited by the parameter $\tau$ value. Even taking this into account, we have shown that our algorithms are able to find optimal solutions for several problem instances (Figs. \ref{fig: capacity-time} and \ref{fig: clases-time}).

Future research will focus on analyzing the impact of the parameter $\tau$ more systematically and developing alternative implementations that avoid numerical overflows before reaching the optimal solution. Additionally, we aim to extend this approach to other combinatorial optimization problems, broadening the applicability of quantum-inspired tensor network methods in this domain, and study its possible implementation in quantum hardware to improve its performance.

\section*{\uppercase{Data and code availability}}
All data and code required for this project can be accessed upon reasonable request by contacting the authors.

\section*{\uppercase{Acknowledgment}}
The research has been funded by the Ministry of Science and Innovation and CDTI under ECOSISTEMAS DE INNOVACIÓN project ECO-20241017 (EIFEDE) and ICECyL (Junta de Castilla y León) under project CCTT5/23/BU/0002 (QUANTUMCRIP). This project has been funded by the Spanish Ministry of Science, Innovation and Universities under the project PID2023-149511OB-I00.

\bibliographystyle{apalike}
{\small
\bibliography{example}}

\onecolumn\newpage
\appendix
\section{\uppercase{Notation for logical tensors}}
The definition of logical tensors can be complicated and unreadable, especially in sparse cases. We have therefore taken a simpler notation to express them.

Take a 4-index $T$ tensor with dimensions $d_0$, $d_1$, $d_2$ and $d_3$, which has only non-zero elements given by a function of the indices, at the positions given by another function of one or more indices. This tensor can be expressed in four blocks of information: the name, the indices, the auxiliaries and the values.

The first thing we will do is to express the dimensions of the tensor in its name, as subscript of products of dimensions. For example, in this tensor its name is
\begin{equation}
    T_{d_0\times d_1\times d_2\times d_3}.
\end{equation}

Each value of the subscripts indicates the dimension of the index of the corresponding position. For when we define the elements of the tensor, we will not add extra indices, but replace these subscripts by the indices of the tensor.

We differentiate the independent indices, which run through all the values of its dimension, from the dependent indices, whose values are obtained from the first ones. The independent ones will be expressed in roman letters, while the dependent ones will be expressed by means of greek letters. The general way of expressing them is
\begin{equation}
    \begin{gathered}
        \mu = f_\mu(i,j),\\
        \nu = f_\nu(i,j).
    \end{gathered}
\end{equation}

If we had that the tensor has an input value $n$ which creates an internal auxiliary variable $y$ to be introduced to the functions, it can be added as
\begin{equation}
    \begin{gathered}
        y = g_y(n),\\
        \mu = f_\mu(i,j),\\
        \nu = f_\nu(i,j).
    \end{gathered}
\end{equation}

The non-zero elements of the tensor will be those whose indices satisfy the above equations, and will be obtained by a series of functions. An example is 
\begin{equation}
    T_{ij\mu\nu} =
    \begin{cases} 
    t_1(i,j,\mu,\nu)&\text{ if }  h_1(i,j,\mu,\nu)>0,\\
    t_2(i,j,\mu,\nu)&\text{ if }  h_2(i,j,\mu,\nu)>0,\\
    t_3(i,j,\mu,\nu)&\text{ else }.
   \end{cases}
\end{equation}

Thus, a tensor in general will be expressed as
\begin{align*}
    &T_{d_0\times d_1\times d_2\times d_3\times \dots\times d_N}\\
    &\begin{gathered}
        y = g_y(n,m,\dots),\\
        \vdots\\
        \mu = f_\mu(i,j,\dots),\\
        \nu = f_\nu(i,j,\dots),\\
        \vdots
    \end{gathered}\\
    T_{ij\dots \mu\nu\dots} =&
    \begin{cases} 
    t_1(i,j,\dots,\mu,\nu,\dots)&\text{ if }  h_1(i,j,\dots,\mu,\nu,\dots)>0,\\
    t_2(i,j,\dots,\mu,\nu,\dots)&\text{ if }  h_2(i,j,\dots,\mu,\nu,\dots)>0,\\
    \vdots\\
    t_M(i,j,\dots,\mu,\nu,\dots)&\text{ else }.
   \end{cases}
\end{align*}

A particular case is a tensor $T^m$ of $N$ indexes in which the non-zero elements are in the elements where the rule is
\begin{align*}
    y_j& = m^j\\
    i_n& = \sum_{j} \left(\omega_j i_j -y_j\right) 
\end{align*}
and its elements have the value
\begin{equation}
        T^m_{i_0,i_1,i_2,\dots,i_{N-1}} = 
    \begin{cases}
        \sum_{j} e^{\beta i_j+m} \text{ if } \sum_j i_j -50 > 0,\\
        \sum_{j} e^{\beta i_j-m} \text{ if } \sum_j i_j -100 > 0,\\
        1.
    \end{cases}
\end{equation}

This tensor can be expressed as
\begin{align*}
    &T^m_{d_0\times d_1\times d_2\times d_3\times \dots\times d_N}\\
    &\begin{gathered}
        y_j = m^j\\
        i_n = \sum_{j} \left(\omega_j i_j -y_j\right)
    \end{gathered}\\
     T^m_{i_0,i_1,i_2,\dots,i_{N-1}} =& 
    \begin{cases}
        \sum_{j} e^{\beta i_j+m} \text{ if } \sum_j i_j -50 > 0,\\
        \sum_{j} e^{\beta i_j-m} \text{ if } \sum_j i_j -100 > 0,\\
        1.
    \end{cases}
\end{align*}

\end{document}